\documentclass[12pt]{article}
\usepackage[utf8]{inputenc}
\usepackage{amsmath, mathtools, amsfonts, amssymb, amsthm, bm, float, subcaption}
\usepackage[shortlabels]{enumitem}
\usepackage[linesnumbered, ruled]{algorithm2e}
\usepackage{color}
\usepackage[title]{appendix}
\usepackage{tikz}
\usetikzlibrary{positioning}
\usepackage{hyperref}
\usepackage{url}
\usepackage{mathtools}
\usepackage{multicol}
\usepackage{setspace}
\usepackage{verbatim}
\usepackage{natbib} 
\usepackage[legalpaper, left=0.75in, right=0.75in, top=1in, bottom=1in]{geometry}
\usepackage{newtxtext}
\usepackage{newtxmath}
\usepackage{natbib}
\usepackage{hyperref}
\hypersetup{
    colorlinks = true,
    urlcolor   = blue,
    citecolor  = black,
}

\newcommand{\RomanNumeralCaps}[1]
\linenumbers

\usepackage{comment}
\usepackage[justification=raggedright,  width=\textwidth]{caption}
\title{\LARGE Phase-based analysis and control of \\ low Reynolds number aeroelastic flows}

\author{
    \normalsize Chathura R. Sumanasiri\textsuperscript{1}\\
    \small \textit{Department of Mechanical Engineering}\\
    \small \textit{University of Nevada, Reno}\\
    \small Reno, USA\\
    \small csumanasiri@unr.edu
    \and
    \normalsize Tulsi Ram Sahu\textsuperscript{2}\\
    \small \textit{Department of Mechanical Engineering}\\
    \small \textit{University of Nevada, Reno}\\
    \small Reno, USA\\
    \small tsahu@unr.edu
    \and
    \normalsize Aditya G. Nair\textsuperscript{3}\\
    \small \textit{Department of Mechanical Engineering}\\
    \small \textit{University of Nevada, Reno}\\
    \small Reno, USA\\
    \small adityan@unr.edu
}

\usepackage{titlesec}

\date{}
\begin{document}
% Redefine paragraph to behave like a block heading
\titleformat{\paragraph}
  {\normalfont\normalsize\bfseries}{\theparagraph}{1em}{} % Format like subsubsection
\titlespacing*{\paragraph}
  {0pt}{3.25ex plus 1ex minus .2ex}{1.5ex plus .2ex} % Add spacing before/after

\onehalfspacing  % Set line spacing to 1.5 throughout the document

\maketitle

\begin{abstract}
\normalsize
Flutter in lightweight airfoils under unsteady flows presents a critical challenge in aeroelastic stability and control. This study uncovers phase-localized mechanisms that drive the onset and suppression of flutter in a freely pitching airfoil at low Reynolds number. By introducing targeted impulsive stiffness perturbations, we identify critical phases that trigger instability. Using phase-sensitivity functions, energy-transfer metrics, and dynamic mode decomposition, we show that flutter arises from phase lock-on between structural and fluid modes. Leveraging this insight, we design an energy-optimal, phase-based control strategy that applies transient heaving motions to disrupt synchronization and arrest unstable growth. This minimal, time-localized control suppresses subharmonic amplification and restores stable periodic motion.
\end{abstract}

\section{Introduction}
\label{sec:headings}

Aeroelastic flutter is a critical instability in lightweight and flexible structures that interact with unsteady flows, characterized by self-excited limit-cycle oscillations (LCOs) that can compromise structural integrity \citep{dowell2021nonlinear}. Controlling these oscillations is essential to improve aerodynamic performance and prevent undesirable responses \citep{jonsson2019flutter}. However, predicting stability transitions remains challenging due to the nonlinear nature of fluid–structure interactions and their sensitivity to flow separation and vortex shedding \citep{schuster2003computational}.

Classical theories assume thin airfoils and inviscid, attached flows \citep{Theodorsen1935, bisplinghoff2013aeroelasticity}, limiting their applicability in regimes dominated by unsteady separation. In such conditions, laminar separation flutter and stall-induced LCOs can emerge from nonlinear interactions between separated flow and structural motion \citep{Barnes2019, lee1999nonlinear}. Recent advances in flutter prediction include multifidelity models \citep{thelen2020aeroelastic}, frequency-domain identification \citep{simiriotis2023numerical}, and data-driven techniques \citep{hickner2023data, guo2025neural}. Complementary efforts in active flutter suppression employ high-bandwidth actuators and adaptive controllers to stabilize flutter-prone modes and mitigate LCOs \citep{Livne2018, li2011adaptive, chai2021aeroelastic}.

However, the impact of localized structural variations remains poorly understood, particularly in systems exhibiting subcritical bifurcations, where small changes can trigger instability \citep{castravete2008effect, khodaparast2010propagation, beran2017uncertainty}. Energy maps have proven useful for understanding the nonlinear aeroelastic response to gusts \citep{Menon2019, Menon2020}; however, their dependence on steady-state responses from forced simulations across a range of parameters limits their utility for real-time flutter characterization.

To address these limitations, phase reduction offers a low-dimensional tractable framework by modeling unsteady periodic flows as nonlinear oscillators \citep{monga2019phase, taira2018phase}. Recent advances include timing control of vortex shedding dynamics \citep{Nair2021}, adjoint-based formulations for rapid wake synchronization \citep{Godavarthi2023}, and data-driven latent manifold approaches for suppressing transient gust responses \citep{Fukami2024}.

Unlike canonical oscillators, aeroelastic systems can exhibit abrupt transient instabilities that lead to flutter, often before reaching a stable periodic orbit. In this study, we apply phase-based methods to characterize and control the onset of such instabilities. Using two-dimensional incompressible flow over a NACA0015 airfoil at a low Reynolds number, we introduce time-localized perturbations during the free-response oscillation cycle and show that their impact is highly phase-dependent. We demonstrate how this sensitivity can be exploited to suppress flutter through targeted, phase-based control strategies. 

\vspace{-0.15in}
\section{Methodology}
\label{sec:methods}

We perform two-dimensional simulations of the aeroelastic response of a NACA0015 airfoil immersed in an incompressible flow using the immersed boundary projection method (IBPM) \citep{taira2007immersed, goza2017strongly}. The fluid motion is governed by the incompressible Navier--Stokes equations in non-dimensional form:
\begin{equation}
\frac{\partial \mathbf{u}}{\partial t} + \mathbf{u} \cdot \nabla \mathbf{u} = -\nabla p + \frac{1}{Re} \nabla^2 \mathbf{u}, \quad \nabla \cdot \mathbf{u} = 0, \label{eq:NS1}
\end{equation}
where $\mathbf{u}$ is the velocity field, $p$ is pressure, and $Re = U_\infty c / \nu$ is the Reynolds number, with $U_\infty$, $c$, and $\nu$ denoting the freestream velocity, chord length, and kinematic viscosity, respectively. IBPM discretizes these equations on a staggered Cartesian grid, employing a Crank–Nicolson scheme for viscous terms and a second-order Adams–Bashforth scheme for convective terms. No-slip boundary conditions are enforced via Lagrange multipliers applied at Lagrangian boundary points, with interpolation and spreading handled via regularized delta functions. A nested multi-domain grid with a fast Poisson solver \citep{colonius2008fast} is used to efficiently resolve near- and far-field interactions.

\begin{figure}
    \begin{center}
    \includegraphics[width=0.95\linewidth]{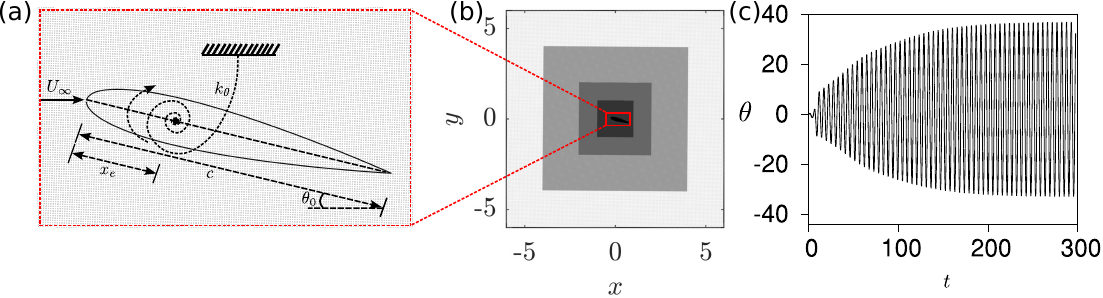}
    \end{center}
    \vspace{-0.1in}
    \caption{
        Free-pitching dynamics of a NACA0015 airfoil. 
        (a) Schematic of the computational setup showing the airfoil mounted at an elastic axis located at $x_e = 0.33c$ from the leading edge. 
        (b) Nested multi-domain mesh layout used for simulations, with finest resolution near the airfoil. 
        (c) Validation of gust-induced limit-cycle oscillations in pitch angle $\theta(t)$, reproducing aeroelastic responses consistent with \citep{Menon2020}.
    }
    \label{fig1}
\end{figure}

The airfoil is free to pitch about an elastic axis located at $x_e$ from the leading edge (Figure~\ref{fig1}a). Its structural dynamics are modeled as a single-degree-of-freedom linear torsional oscillator governed by
\begin{equation}
    I^* \ddot{\theta} + k^*(\theta - \theta_0) = C_M, \label{eq:structure}
\end{equation}
where $\theta$ is the pitch angle, $\theta_0$ is the equilibrium angle, $C_M$ is the dimensionless aerodynamic moment, and $I^*$ and $k^*$ are the non-dimensional moment of inertia and stiffness, respectively.

We adopt parameter values consistent with those used by \citep{Menon2020}, who identified an aeroelastic regime exhibiting strong nonlinear fluid–structure coupling and the onset of flutter. The selected parameters are $Re = 1000$, $x_e = 0.33c$, $\theta_0 = 15^\circ$, $I^* = 4.18$, and $k^* = 5.96$. The structural natural frequency is $f_s = (1/2\pi)\sqrt{k^*/I^*} \approx 0.19$ and the associated time period is $T = 1/f_s$.

The computational domain spans $x/c \in [-16, 16]$ and $y/c \in [-16, 16]$. The finest grid near the airfoil has a uniform spacing $\Delta x = \Delta y = 0.0055c$, and a fixed time step $\Delta t = 0.001$ is used for time integration. The near-body mesh and nested multi-domain layout are shown in Figure~\ref{fig1}(a) and (b), respectively. The setup is validated by reproducing gust-induced LCO responses as seen in Figure~\ref{fig1}(c), consistent with \citep{Menon2020}, confirming the nonlinear amplification behavior near the onset of flutter.

\begin{figure}
    \centering
    \includegraphics[width=1\linewidth]{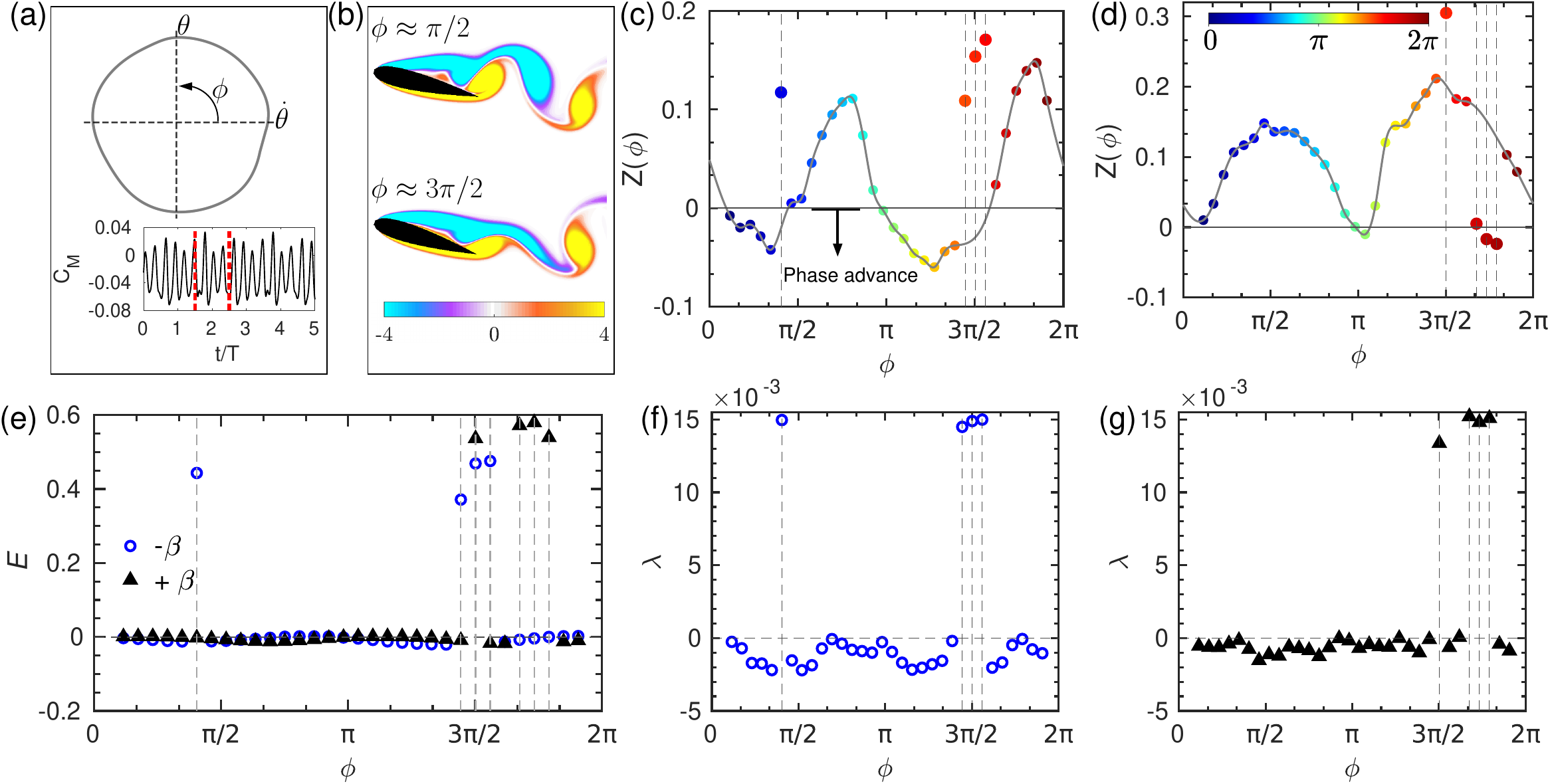}
    \vspace{-0.2in}
    \caption{
Phase-based analysis of aeroelastic response to impulsive stiffness perturbations. 
(a) Definition of phase $\phi$ on the $\dot{\theta}$–$\theta$ plane (top) and time series of the aerodynamic moment coefficient $C_M$ (bottom), with red vertical lines indicating one period of pitch oscillation. 
(b) Instantaneous vorticity fields at representative phases.
(c,d) Phase-sensitivity function $Z(\phi)$ estimated from asymptotic pitch angle shifts following negative (decrease in $k^*$) and positive (increase in $k^*$) stiffness impulses. 
Dashed lines mark phases where perturbations lead to divergent flutter-like behavior. 
(e) Normalized energy extraction $E$ computed over 18 post-perturbation cycles, showing increased energy transfer at critical phases. 
(f,g) Exponential growth rate $\lambda$ of the envelope of $C_M(t)$, with positive values indicating instability. 
}
    \label{fig:PS}
\end{figure}

To investigate the timing-dependent sensitivity of the aeroelastic system, we adopt the observable-based phase-reduction framework of \citep{taira2018phase}, which projects the high-dimensional fluid--structure dynamics onto a scalar phase variable defined along the periodic orbit. Specifically, we define the phase \( \phi \) on the \( \dot{\theta} \)--\( \theta \) plane, where \( \theta \) denotes the airfoil pitch angle and \( \dot{\theta} \) is the angular velocity. This observable-based representation traces the aeroelastic limit cycle with period \( T \) and provides a smooth, monotonically increasing phase coordinate over the oscillation cycle, as shown in the polar plot of Figure~\ref{fig:PS}(a, top). The bottom panel of Figure~\ref{fig:PS}(a) shows the corresponding time-series of the aerodynamic moment coefficient \( C_M \), with vertical red lines marking the bounds of one period of the pitch oscillation $T$. Within this interval, \( C_M \) exhibits higher-frequency oscillations compared to \( \theta \), with a dominant frequency of \( f \approx 0.67 \), which is also observed when the airfoil is held static. This frequency mismatch between \( \theta \) and \( C_M \) in the free-response indicates a lack of synchronization between structural motion and aerodynamic forcing, thereby preventing sustained energy transfer and averting flutter in the baseline configuration. Figure~\ref{fig:PS}(b) presents representative vorticity fields at two characteristic phases, revealing the evolution of coherent vortical structures around the airfoil as it traverses the limit cycle.

Under an external perturbation applied at time $t_0$, the phase dynamics evolves as
\begin{equation}
\dot{\phi}(t) = 2\pi f_s + \epsilon Z(\phi)\delta(t - t_0),
\end{equation}
where $Z(\phi)$ is the phase-sensitivity function (PSF), $\epsilon$ is the perturbation magnitude, and $\delta$ is the Dirac delta function centered at the impulse time $t_0$ \citep{nakao2016phase}. Following the impulse-response method of \citep{taira2018phase}, we apply Gaussian-shaped impulses at uniformly-spaced phases across the oscillation cycle as 
\begin{equation}
\epsilon \delta(t-t_0) \approx \frac{\beta}{\sqrt{2\pi\sigma}} \exp\left[-\frac{1}{2} \left( \frac{t - t_0}{\sigma} \right)^2 \right],
\end{equation}
where $\beta$ and $\sigma = 0.0019T$ are the strength and width of the impulse introduced at time \( t_0 \), respectively. These impulses are introduced by perturbing structural parameters (e.g., stiffness) or by imposing surging/heaving velocity impulses on the airfoil. In this study, stiffness perturbations are used to probe phase-based flutter sensitivity due to their direct relevance to structural degradation under cyclic loading. Stiffness variations, arising naturally from material fatigue or geometric non-uniformities, directly affect the system's natural frequency $f_s$, making them effective triggers for phase-dependent stability transitions.

The phase change induced by each perturbation is evaluated relative to the unperturbed baseline by tracking the observable $\theta(t)$. The phase-sensitivity function (PSF) is then approximated as $Z(\phi) \approx \frac{\Delta \phi}{\beta}$, where $\Delta \phi$ denotes the asymptotic phase difference between the baseline and the perturbed trajectories. A positive value of $Z(\phi)$ implies a phase delay (i.e., the perturbed trajectory lags the baseline), while a negative value indicates a phase advance. This observable-based formulation bypasses the need for full-state measurements and provides an experimentally feasible way to quantify phase sensitivity in high-dimensional aeroelastic systems.

Once $Z(\phi)$ is known, we design an optimal open-loop control input $u(t)$ to advance or delay the oscillation phase over a finite time horizon. This strategy leverages the observable-based phase-reduction model to implement timing-sensitive actuation without requiring full-state feedback. The controlled phase dynamics is
\begin{equation}
\dot{\phi}(t) = 2\pi f_s + u(t) Z(\phi),
\label{eq:control}
\end{equation}
with the objective of achieving a prescribed phase shift $\Delta \phi$ while minimizing control effort. This leads to a two-point boundary value problem, which can be solved using standard calculus of variations techniques \citep{monga2019phase, Nair2021}. We assess the effectiveness of this strategy in suppressing the onset of flutter-like instability.

\vspace{-0.15in}
\section{Results}

We first examine the effect of impulsive stiffness perturbations across the free-response oscillation cycle. Gaussian-shaped impulses are applied to the stiffness parameter $k^*$ in Eq.~(\ref{eq:structure}) at 32 uniformly spaced phases, and the resulting system evolution is monitored. Figure~\ref{fig:PS}(c) and (d) shows the estimated phase-sensitivity function $Z(\phi)$ for negative (decrease in $k^*$ making it less stiff impulsively) and positive (increase in $k^*$) stiffness perturbations, respectively, computed from asymptotic shifts (after 19 fundamental oscillation cycles) in pitch angle trajectories as detailed in \S \ref{sec:methods}.

The sensitivity functions reveal distinct regions of phase advance ($Z(\phi) < 0$) and phase delay ($Z(\phi) > 0$), reflecting how timing of perturbation affects the flow–structure coupling. For negative stiffness perturbations [Figure~\ref{fig:PS}(c)], four specific phases, $\phi \approx 0.353\pi$, $1.410\pi$, $1.472\pi$, and $1.526\pi$, lead to divergent responses and trigger flutter onset. Similarly, in the case of positive perturbations [Figure~\ref{fig:PS}(d)], flutter is initiated when impulses are applied at $\phi \approx 1.472\pi$, $1.656\pi$, $1.712\pi$, and $1.768\pi$. These critical phases are marked with vertical dashed lines and are associated with sharp variations in $Z(\phi)$. The asymmetry between sensitivity functions for positive and negative perturbations stems from stiffness being a material property. Unlike velocity-based actuation, stiffness perturbations directly alter the natural frequency \( f_s \), leading to inherently nonlinear and directionally biased responses. This contrasts with prior studies \citep{taira2018phase}, where momentum-based impulses yielded symmetric \( Z(\phi) \).

The structure of $Z(\phi)$ further reveals distinct phase-dependent regimes that align with the dynamics in the $\dot{\theta}$–$\theta$ phase plane [Figure~\ref{fig:PS}(a)]. In the case of negative stiffness perturbations [Figure~\ref{fig:PS}(c)], four alternating intervals are observed: $Z(\phi) < 0$ from $0$ to $\pi/2$ and again from $\pi$ to $3\pi/2$, indicating phase advance during downstroke and upstroke, respectively; and $Z(\phi) > 0$ from $\pi/2$ to $\pi$ and from $3\pi/2$ to $2\pi$, indicating phase delay near the turning points of pitch motion. These transitions reflect how structural input interacts with unsteady aerodynamic loads depending on the direction of motion. In contrast, the positive perturbation case [Figure~\ref{fig:PS}(d)] displays two broad lobes of phase delay, peaking near $\phi = \pi/2$ and $3\pi/2$, where pitch angle $\theta$ reaches its maxima. Although the precise locations of instability differ, both cases demonstrate that flutter onset is strongly localized in phase. For both perturbations, $Z \rightarrow 0$ at $\phi = 0$ and $\pi$ as $\theta = 0$ at these phases and thus the stiffness perturbation does not modify the structural dynamics of Eq.~\ref{eq:structure} as much.

To evaluate the response of the aeroelastic system following impulsive stiffness perturbations, we compute the normalized energy extraction coefficient $E$ over a transient window starting immediately after the period in which the impulse is applied and spanning 18 fundamental oscillation cycles. This energy metric, adapted from \citep{Menon2020}, is 
\begin{equation}
E = \frac{1}{(T_{f}-T_i)} \int_{T_i}^{T_{f}} C_M(t) \, \dot{\theta}(t) \, dt,
\label{eq:energy}
\end{equation}
where $C_M(t)$ is the aerodynamic moment coefficient, $\dot{\theta}(t)$ is the pitch rate, $T_i = T$ is the initial time, and $T_f = 19T$ is the final time. The sign and magnitude of $E$ indicate whether the fluid imparts energy into the structure (positive $E$) or extracts energy from it (negative $E$) over the transient evolution. As shown in Figure~\ref{fig:PS}(e), negative  stiffness perturbations (circles) and positive perturbations (triangles) yield large positive $E$ values at specific phases, aligning with those identified in Figures~\ref{fig:PS}(c--d) as prone to flutter onset. These high-energy-extraction phases mark the system's transition from stable periodic motion to divergent, flutter-like growth.

To quantify the instability, we fit an exponential envelope to the pitch moment coefficient $C_M(t)$ using $\max(C_M)(t) \sim C_0 e^{\lambda t}$, where $\lambda$ is the instantaneous growth rate. Figures~\ref{fig:PS}(f--g) show $\lambda$ for negative and positive stiffness perturbations, respectively. Positive values of $\lambda$ indicate exponential growth and align with phase regions exhibiting high energy $E$ and sharp variations in $Z(\phi)$, while negative values reflect damping and return to periodic behavior. The consistency across $\lambda$, $Z(\phi)$, and $E$ confirms that flutter onset is phase-localized and can be predicted using phase-based metrics.

\begin{figure}
    \centering
    \includegraphics[width=1\linewidth]{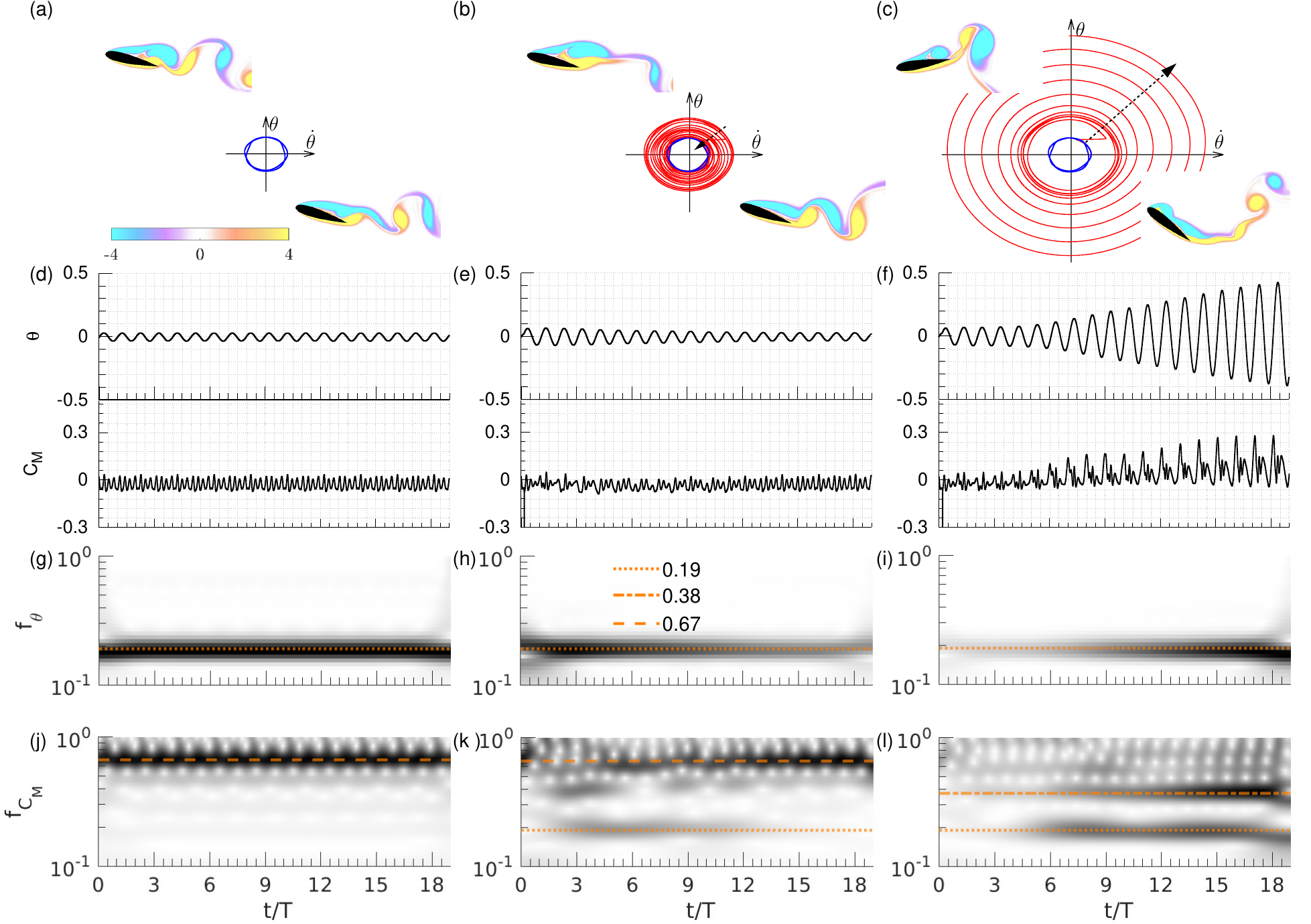}
    \vspace{-0.2in}
    \caption{Transient response to impulsive stiffness perturbations at selected phases. (a,d,g,j) Baseline (unperturbed) response; (b,e,h,k) stable response to perturbation at $\phi = 0.29\pi$; (c,f,i,l) unstable (flutter) response to perturbation at $\phi = 0.353\pi$. (a–c) Phase portraits in the $\dot{\theta}$–$\theta$ plane with unperturbed (blue) and perturbed (red) trajectories, overlaid with instantaneous vorticity fields. (d–f) Time series of pitch angle $\theta(t)$ and aerodynamic moment coefficient $C_M(t)$. (g–i) Time–frequency spectrograms of $\theta(t)$ and (j–l) of $C_M(t)$, computed using continuous wavelet transforms to capture joint temporal and spectral features.}
    \label{fig:lock-in}
\end{figure}

To visualize the transient response to stiffness perturbations, we compare three representative cases in Figure~\ref{fig:lock-in}: (i) the unperturbed baseline (left column), (ii) a stable response to a perturbation applied at $\phi = 0.29\pi$ (middle column), and (iii) an unstable (flutter) response resulting from a perturbation at $\phi = 0.353\pi$ (right column).

The top row (Figures~\ref{fig:lock-in}(a–c)) shows phase portraits in the $\dot{\theta}$–$\theta$ plane, with the baseline trajectory in blue and the perturbed trajectory in red. For the stable case, the perturbed trajectory initially deviates but gradually converges back to the baseline limit cycle, demonstrating recovery to the original periodic motion. In contrast, the flutter case exhibits sustained divergence and expanding amplitude, indicating exponential growth and onset of instability. Overlaid instantaneous vorticity fields at representative instants capture the evolution of the flow structures. In the baseline and stable cases, the flow remains largely periodic with limited fluctuation in vortex strength. However, in the flutter case, the larger structural amplitude leads to intensified vortex shedding and greater flow separation.

The middle row (Figures~\ref{fig:lock-in}(d–f)) presents the corresponding time histories of pitch angle $\theta(t)$ and aerodynamic moment coefficient $C_M(t)$. The stable case reveals bounded oscillations that settle after the transient, whereas the flutter case displays an exponential increase in both $\theta$ and $C_M$, consistent with the positive growth rate $\lambda$ observed in Figure~\ref{fig:PS}(f).

\newpage

 The bottom two rows present time–frequency spectrograms of $\theta(t)$ (Figures~\ref{fig:lock-in}(g–i)) and $C_M(t)$ (Figures~\ref{fig:lock-in}(j–l)), obtained using continuous wavelet transforms to capture both temporal and spectral dynamics.  All three cases exhibit a consistent dominance of structural natural frequency $f_s$ in $\theta$, validating the robustness of the structural response. However, the spectrograms of $C_M$ provide more nuanced insights into the flow-structure interaction. For the baseline case, $C_M$ exhibits a persistent high-frequency content around $f_{C_M} \approx 0.67$, associated with vortex shedding dynamics. In the stable perturbed case, a transient amplification of low-frequency components at $f_{C_M} \approx 0.19$ (matching $f_s$) and its harmonic at $f_{C_M} \approx 0.38$ is observed, but these decay over time. In contrast, the flutter case shows a continuous and growing amplification of these low-frequency components post-perturbation, indicating sustained energy transfer from the flow to the structure. This progressive emergence and amplification of the dominant instability $f_{C_M} \approx 0.38$ and its subharmonic $f_{C_M} \approx 0.19$ (matching $f_s$) in $C_M$ reflect a phase lock-on phenomenon between the aerodynamic forcing and the structural motion. This lock-on effectively channels energy into the structural mode, resulting in flutter. 

To further examine the underlying flow structures contributing to stability and instability, we apply dynamic mode decomposition (DMD) \citep{schmid2022dynamic} to the unsteady flow fields following impulsive stiffness perturbations. Specifically, we compare the flutter-prone unstable case ($\phi = 0.353\pi$) and the stable case ($\phi = 0.29\pi$) discussed in Figure~\ref{fig:lock-in}. DMD is performed in a body-fixed frame to isolate flow oscillations associated with the airfoil's pitching motion.

DMD approximates the system’s evolution using a linear operator $A$ such that $\mathbf{\omega}_{k+1} \approx A \mathbf{\omega}_k$, where $\mathbf{\omega}_k$ denotes the vorticity field at time $t_k$. The discrete-time eigenvalues $\lambda_j$ of $A$ encode the growth rate and frequency of the corresponding DMD modes $\Phi_j$. The continuous time spectra are obtained as $\Lambda_j = \frac{\log(\lambda_j)}{\Delta t}$, where $\Re(\Lambda_j)$ gives the modal growth rate and $f_\Phi = \Im(\Lambda_j)/2\pi$ gives the oscillation frequency. The mode amplitudes $b_j$ are computed by projecting the initial state $\mathbf{\omega}_0$ onto the DMD modes via $\mathbf{b} = \Phi^\dagger \mathbf{x}_0$, where $\dagger$ denotes the Moore--Penrose pseudoinverse.

\begin{figure}
    \centering
    \includegraphics[width=0.95\linewidth]{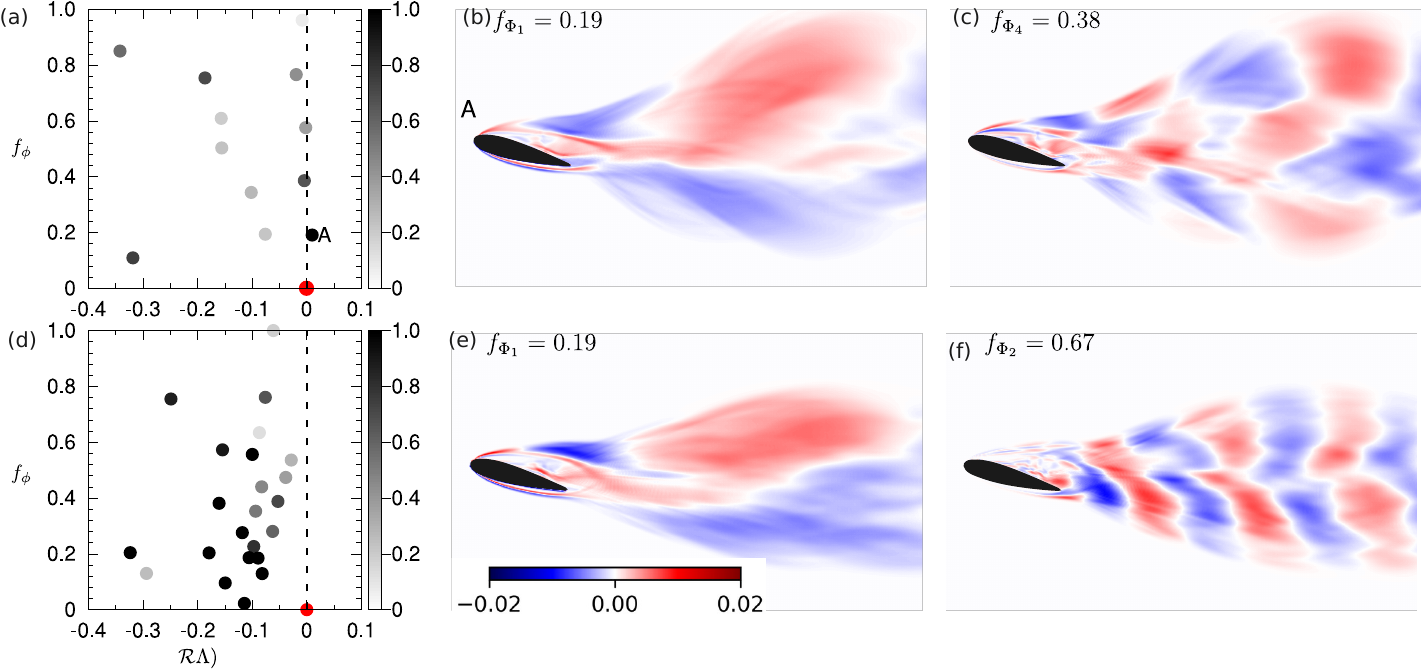}
    \vspace{-0.1in}
    \caption{Dynamic Mode Decomposition (DMD) of the flow field for stiffness perturbations at $\phi = 0.353\pi$ (flutter-prone, top row) and $\phi = 0.29\pi$ (stable, bottom row). (a, d) DMD eigenvalues plotted in terms of growth rate versus frequency, with colors indicating mode amplitudes $|b|$. (b, e) DMD spatial mode at $f_\Phi \approx 0.19$ (structural frequency). (c, f) DMD spatial mode at $f_\Phi \approx 0.38$ (first harmonic) for flutter-prone case and spatial mode at $f \approx 0.67$ for stable case. DMD is performed in the lab frame aligned with airfoil pitching motion. In panels (a) and (d), the red dot indicates the DMD mean mode.}
    \label{fig:dmd}
\end{figure}

The top row of Figure~\ref{fig:dmd} shows results for the flutter-prone unstable case. Panel (a) presents the DMD eigenvalues projected in the complex plane, with the growth rate $\Re(\Lambda_j)$ on the $x$-axis and frequency $f_\Phi$ on the $y$-axis. The markers are colored by modal amplitude $|b_j|$. The dominant modes appear at $f_\Phi \approx 0.38$ and its subharmonic $f_\Phi \approx 0.19$. The subharmonic mode with $f_\Phi \approx 0.19$ has a positive growth rate, which confirms it as the driver of flutter. The corresponding spatial structures, shown in Figures~\ref{fig:dmd}(b) and (c), exhibit strong vortex shedding from the trailing edge and large-scale separation from the suction surface. The bottom row presents results for the stable case. In Figure~\ref{fig:dmd}(d), all eigenvalues lie in the left-half complex plane, confirming that the dynamics are stable. The dominant modes, shown in Figures~\ref{fig:dmd}(e) and (f), correspond to frequencies $f_\Phi \approx 0.19$ and $f_\Phi \approx 0.67$ (associated with moment coefficient oscillations). These findings support the conclusion that flutter is driven by phase lock-on between structural and fluid modes, resulting in selective energy amplification of low-frequency DMD modes. In the stable case, this coupling remains limited, preventing instability.

To suppress the flutter-prone instability triggered by negative stiffness perturbation at \( \phi = 0.353\pi \), we introduce a phase-based control (PBC) strategy through transient heaving motion of the airfoil. Although the instability is induced via stiffness perturbations, stiffness itself is not easily actuated in practice. Heaving motion, on the other hand, can be implemented more readily through control surfaces, making it a physically realizable control input. 

\begin{figure}
    \centering
    \includegraphics[width=1\linewidth]{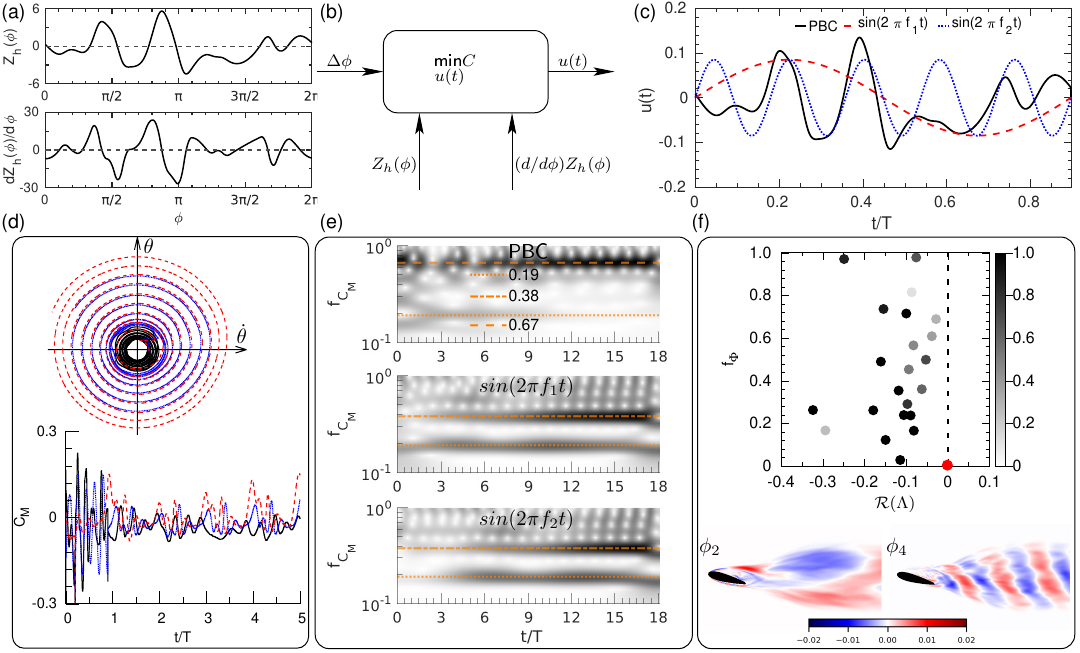}
    \vspace{-0.2in}
    \caption{Phase-based control (PBC) of flutter instability via heaving actuation. (a) Phase-sensitivity function and its gradient for heaving input. (b) Schematic of the energy-optimal phase-control framework targeting a prescribed phase shift. (c) Control inputs: optimal phase-based actuation (black), low-frequency sinusoid ($f_1 = f_s/0.9$), and high-frequency sinusoid ($f_2 = 1.05$). (d) Resulting pitch angle $\theta(t)$ and moment coefficient $C_M(t)$. (e) Spectrograms of $C_M(t)$ for each actuation. (f) DMD eigenvalues and dominant modes for the phase-based control case.}
    \label{fig:control}
\end{figure}

Following the phase-based control framework described in \S\ref{sec:methods}, we compute the phase-sensitivity function \( Z_h(\phi) \) and its gradient \( \frac{dZ_h}{d\phi} \) by applying Gaussian-shaped (downward) heaving impulses at different phases of the oscillation cycle with impulse strength \( \beta = -0.01 \). The resulting sensitivity profile, shown in Figure~\ref{fig:control}(a), reveals alternating regions of phase advance and delay, identifying windows of high control effectiveness.

With this phase sensitivity, we formulate an control problem to compute an energy-optimal input \( u(t) \) that induces a prescribed phase shift \( \Delta \phi \) over a finite time horizon. The underlying rationale is that by advancing the phase trajectory to reach the same oscillatory position slightly ahead of its natural timing, we disrupt the temporal coherence between structural motion and unsteady aerodynamic forcing. This intentional desynchronization weakens the feedback loop responsible for energy transfer and flutter amplification. The control input is obtained by minimizing the cost function:
\begin{equation}
    \mathcal{C} = \min_{u(t)} \quad \int_0^{T^*} \left[ \frac{1}{2} u^2(t) + \lambda(t) \left( \dot{\phi}(t) - 2\pi f_s - Z_h(\phi) u(t) \right) \right] dt,
    \label{eq:control}
\end{equation}
where \( \lambda(t) \) is the Lagrange multiplier enforcing the phase evolution constraint, and \( T^* = 0.9T \) is the reduced target period. The resulting optimal control signal \( u(t) \), shown in Figure~\ref{fig:control}(c), is localized in time and requires minimal energy. While the present approach targets phase manipulation, in theory, the same framework could be extended to suppress specific unstable modes by directly shaping the modal energy growth through adjoint-informed or data-driven controllers.

For comparison, we also consider two sinusoidal reference actuations with identical input power as PBC: one with frequency $f_1 = f_s / 0.9$ (shown in red) and the other with $f_2 = 1.05$ (shown in blue). Here, $f_2$ is obtained by selecting the dominant frequency of PBC. The heaving actuation is turned off after $T^*$, allowing the system to evolve freely. We note that negative control input refers to downward heave motion of the airfoil. The corresponding pitch angle and moment coefficient responses are shown in Figure~\ref{fig:control}(d). Only the phase-based control successfully stabilizes the dynamics and suppresses flutter growth. In contrast, the low-frequency sinusoid accelerates divergence, while the high-frequency sinusoid delays instability but fails to prevent it.

Spectrograms of $C_M(t)$ [Figure~\ref{fig:control}(e)] reveal that only the phase-based controller suppresses the subharmonic growth at $f \approx 0.19$, while the sinusoidal inputs result in lock-on to the structural frequency. This underscores the importance of actuation timing and supports the role of phase-specific control in interrupting energy transfer to unstable modes. DMD analysis of the vorticity field under phase-based control [Figure~\ref{fig:control}(f)] shows all modal eigenvalues (with the dominant spatial modes shown below) in the left-half complex plane, indicating fully stabilized dynamics.

\vspace{-0.15in}

\section{Conclusion}

This study demonstrates a phase-based framework for characterizing and controlling flutter instabilities in a freely-pitching airfoil. By applying localized stiffness perturbations across oscillation phases, we uncover critical phase regions that trigger instability, supported by transient energy growth, exponential amplification, and spectral signatures. Dynamic Mode Decomposition further reveals low-frequency mode lock-on as the underlying mechanism of flutter onset. Building on these insights, we design an energy-optimal phase control strategy using transient heaving motion. The proposed controller successfully suppresses flutter by preventing phase lock-on and inhibiting subharmonic amplification. Our findings highlight the critical role of phase in nonlinear aeroelastic interactions and for controlling flow-induced instabilities. 
This phase-based framework can be extended to investigate a wide range of flow-induced phenomena across large parameter spaces, including gust-driven and transonic flutter and aeroacoustic instabilities. While the control strategy is demonstrated at low Reynolds numbers, it holds promise for implementation in supersonic and turbulent regimes \citep{godavarthi2025phase}.

\section*{Acknowledgements}
A.G. Nair acknowledges support from the Air Force Office of Scientific Research (Award No. FA9550-23-1-0483; Program Manager: Dr. Gregg Abate). 

\section*{Data and Code Availability}
The data and code that support the findings are openly available on \href{https://github.com/chathu14/Phase-based-control-for-low-Re-number-flows}{GitHub}.

% Include the bibliography
\bibliographystyle{unsrt}
\bibliography{ref}
\end{document}